\documentclass{article}

\usepackage{fullpage}
\usepackage{amsmath}
\usepackage{amssymb}
\usepackage{derivative}
\usepackage{float}
\usepackage{array}
\usepackage{graphicx}
\usepackage[svgnames]{xcolor}
\usepackage[most]{tcolorbox}
\usepackage[colorlinks=true,bookmarksnumbered,citecolor=DarkGreen,linkcolor=DarkRed]{hyperref}
\usepackage[nameinlink]{cleveref}

\newcolumntype{x}[1]{>{\centering\arraybackslash\hspace{0pt}}p{#1}}
\newcommand{\mse}{\mathsf{mse}}

\makeatletter
\let \@sverbatim \@verbatim
\def \@verbatim {\@sverbatim \verbatimplus}
{\catcode`'=13 \gdef \verbatimplus{\catcode`'=13 \chardef '=13 }} 
\makeatother

\begin{document}

\title{\textbf{Neural Networks, Inside Out:\\Solving for Inputs Given Parameters}\\(A Preliminary Investigation)}
\author{Mohammad Sadeq Dousti\\[1ex] \footnotesize{Institute of Computer Science, Johannes Gutenberg University Mainz, Germany}}
\date{}
\maketitle

\begin{abstract}
\noindent
Artificial neural network (ANN) is a supervised learning algorithm, where parameters are learned by several back-and-forth iterations of passing the inputs through the network, comparing the output with the expected labels, and correcting the parameters. Inspired by a recent work of Boer and Kramer (2020), we investigate a different problem: Suppose an observer can view how the ANN parameters evolve over many iterations, but the dataset is oblivious to him. For instance, this can be an adversary eavesdropping on a multi-party computation of an ANN parameters (where intermediate parameters are leaked).
Can he form a system of equations, and solve it to recover the dataset?\\

\end{abstract}

\section{Introduction}
In artificial neural network (ANN), a labeled dataset is used to train the model with multiple layers of neurons. Each neuron in layer $i$ is connected to one or more neurons in layer $i+1$. Each neuron takes its inputs from the ones connected to it in the previous layer, and computes a function involving the input values and some parameters. The function is usually a liner combination, after which a non-liner function such as the \emph{hyperbolic tangent} $\tanh(\cdot)$ or \emph{rectified linear unit} $\mathrm{ReLU}(\cdot)$ is applied. Initially, the parameters are chosen randomly. Then, in a forward-propagation step, the dataset is given at the input layer, and the value of each neuron is computed until the output layer. Thereafter, the result is compared to the actual labels in the dataset. A loss function, such as \emph{mean squared error} (MSE) shows how distant the computed and real labels are. Then, a procedure such as gradient descent is applied, in order to correct the parameters in a back-propagation step. The whole process is repeated in a number of iterations of epochs.

In a recent work of Boer and Kramer \cite{boer2020secure}, the problem of private learning the parameters of an ANN by multiple parties is considered. That is, the dataset is assumed to be horizontally partitioned%
\footnote{In horizontal partitioning, each party has a subset of rows in the dataset. Each row includes all features as well as the label.}
among multiple parties, and they want to learn the ANN without leaking their private inputs. This can be useful, for instance, when multiple hospitals want to collaboratively learn an ANN on the patient data, while not leaking their inputs to the others, due to privacy concerns.

The main focus of that work is preserving both privacy and efficiency, and therefore the authors did not use encryption. Rather, they cleverly used local computations and a \emph{secure sum protocol} to perform the back-propagation step. However, the updated parameters in each epoch is broadcasted to all parties. Therefore, their protocol is not a secure multiparty computation in the classical sense. (In that sense, no party should observe any intermediate result.)

In this work, we investigate the issue of how harmful their protocol is with respect to privacy. To put it more precisely, we investigate whether an adversary can obtain the \emph{exact} values of the parties' inputs, given he observes the way the parameters of an ANN evolve. The emphasis here is on the exactness, and we do not target partial knowledge gained by such a protocol, which is of independent interest.

To the best of our knowledge, this is the first attempt at solving an ANN for inputs, given the parameters at each epoch. As ordinarily, we solve ANNs for parameters given a set of inputs, we coin the term ``ANNs inside out'' for the problem we are going to solve.

The rest of this paper is as follows: \Cref{sec:minimal} considers the problem for the simplest case: An ANN with just one input and one output case. It is shown how a (non-linear) system of equations can be derived, and the difficulties in solving such a system even in this simple case. \Cref{sec:general} generalizes the ANN to multiple nodes and layers, and also considers ANNs that incorporate randomized learning. Finally, \Cref{sec:conclusion} concludes the paper. In \Cref{sec:apx}, we list the Python code used for generating some tables in the paper.

\section{The simplest case: A minimal neural network}	\label{sec:minimal}

\begin{figure}[H]
\centering
\includegraphics[width=10cm]{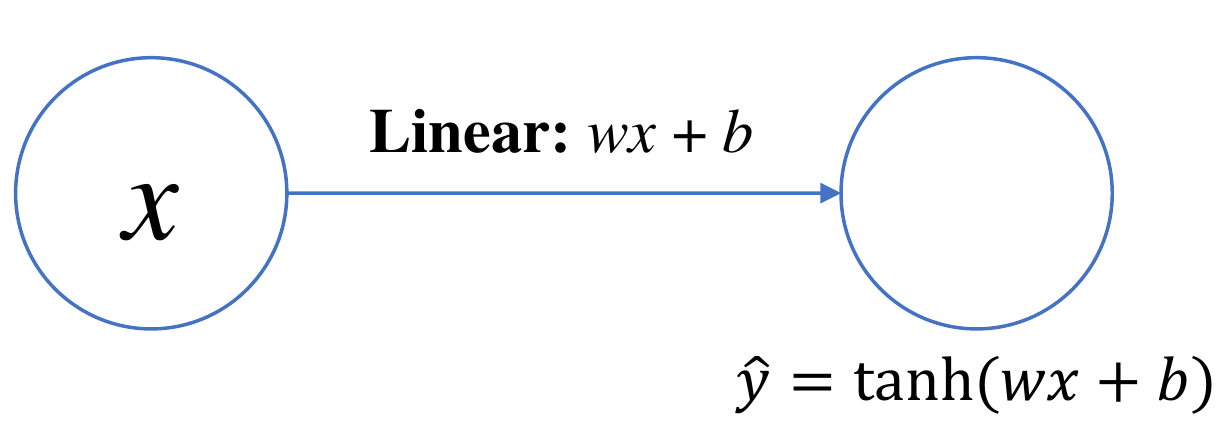}
\caption{Simplest neural network.}
\label{fig:nn1}
\end{figure}

Let us consider the simplest neural network with with 1 input and 1 output, as depicted in \Cref{fig:nn1}. The network has two parameters $w$ and $b$, which are evolved over several epochs. Their values at epoch $j$ is denoted by $w^{(j)}$ and $b^{(j)}$, respectively.

Let the input and output instance vectors be $\mathbf{x} = (x_1, \ldots, x_n)$ and $\mathbf{y} = (y_1, \ldots, y_n)$, respectively.
Assuming the loss function is MSE, and gradient descent is used (note the identity $\odv{}{z}\tanh(z) = 1 - \tanh^2(z)$):

\begin{align}
	\hat{y}_i^{(j)} &= \tanh(w^{(j)}x_i + b^{(j)}) \\
	\mse^{(j)} &= \frac{1}{n}\sum_{i=0}^{n-1} (\hat{y}_i^{(j)} - y_i)^2 \\
	\pdv{\mse^{(j)}}{b^{(j)}} &= \frac{1}{n}\sum_{i=0}^{n-1} 2(\hat{y}_i^{(j)} - y_i)(1 - {(\hat{y}_i^{(j)})}^2) \\
	\pdv{\mse^{(j)}}{w^{(j)}} &= \frac{1}{n}\sum_{i=0}^{n-1} 2x_i(\hat{y}_i^{(j)} - y_i)(1 - {(\hat{y}_i^{(j)})}^2)
\end{align}
The initial weight and bias ($w^{(0)}$ and $b^{(0)}$) are picked uniformly from $[0,1]$.
At each epoch, the weight and bias are updated by the following formulas, where $\eta$ is the learning rate:
\begin{align}
	w^{(j+1)} &= w^{(j)} - \eta \pdv{\mse^{(j)}}{w^{(j)}} \\
	b^{(j+1)} &= b^{(j)} - \eta \pdv{\mse^{(j)}}{b^{(j)}}
\end{align}

\subsection{Numerical examples}
In this section, we assume that $\eta=0.1$ and initially, $w=b=0.5$. We then trace the computations for five epochs. The code used to generate the values in these tables can be found in \Cref{sec:apx}.

\begin{table}[H]
\centering
\caption{$\mathbf{x} = (0.6)$ and $\mathbf{y} = (0.5)$.}
\label{tbl:1}
\begin{tabular}{|c|x{14ex}|x{14ex}|x{14ex}|x{14ex}|x{14ex}|}
	\hline
	epoch & 0 & 1 & 2 & 3 & 4\\
	\hline
	$w^{(j)}$ & 0.500 & 0.489 & 0.479 & 0.469 & 0.460 \\
	\hline
	$b^{(j)}$ & 0.500 & 0.482 & 0.464 & 0.448 & 0.433 \\
	\hline
	$\hat y^{(j)}$ & 0.664 & 0.650 & 0.636 & 0.623 & 0.610 \\
	\hline
	$\mathsf{loss}^{(j)}$ & 0.027 & 0.022 & 0.019 & 0.015 & 0.012 \\
	\hline
\end{tabular}
\end{table}

\begin{table}[H]
\centering
\caption{$\mathbf{x} = (0.6, 0.2)$ and $\mathbf{y} = (0.5, 0.4)$}
\label{tbl:2}
\begin{tabular}{|c|x{14ex}|x{14ex}|x{14ex}|x{14ex}|x{14ex}|}
	\hline
	epoch & 0 & 1 & 2 & 3 & 4\\
	\hline
	$w^{(j)}$ & 0.500 & 0.493 & 0.486 & 0.479 & 0.473 \\
	\hline
	$b^{(j)}$ & 0.500 & 0.481 & 0.463 & 0.447 & 0.432 \\
	\hline
	$\hat y^{(j)}$ & (0.664, 0.537) & (0.651, 0.522) & (0.638, 0.508) & (0.626, 0.495) & (0.615, 0.483) \\
	\hline
	$\mathsf{loss}^{(j)}$ & 0.023 & 0.019 & 0.015 & 0.012 & 0.010 \\
	\hline
\end{tabular}
\end{table}

\begin{table}[H]
\centering
\caption{$\mathbf{x} = (0.6, 0.2, 0.1)$ and $\mathbf{y} = (0.5, 0.4, 0.3)$.}
\label{tbl:3}
\begin{tabular}{|c|x{14ex}|x{14ex}|x{14ex}|x{14ex}|x{14ex}|}
	\hline
	epoch & 0 & 1 & 2 & 3 & 4\\
	\hline
	$w^{(j)}$ & 0.500 & 0.494 & 0.488 & 0.483 & 0.479 \\
	\hline
	$b^{(j)}$ & 0.500 & 0.477 & 0.456 & 0.437 & 0.420 \\
	\hline
	$\hat y^{(j)}$ & (0.664, 0.537, 0.501) & (0.649, 0.520, 0.483) & (0.635, 0.504, 0.466) & (0.621, 0.488, 0.451) & (0.609, 0.474, 0.436) \\
	\hline
	$\mathsf{loss}^{(j)}$ & 0.029 & 0.023 & 0.019 & 0.015 & 0.012 \\
	\hline
\end{tabular}
\end{table}

\begin{table}[H]
\centering
\caption{$\mathbf{x} = (0.6, 0.2, 0.1, 0.9)$ and $\mathbf{y} = (0.5, 0.4, 0.3, 0.6)$.}
\label{tbl:4}
\begin{tabular}{|c|x{14ex}|x{14ex}|x{14ex}|x{14ex}|x{14ex}|}
	\hline
	epoch & 0 & 1 & 2 & 3 & 4\\
	\hline
	$w^{(j)}$ & 0.500 & 0.493 & 0.486 & 0.479 & 0.473 \\
	\hline
	$b^{(j)}$ & 0.500 & 0.480 & 0.461 & 0.444 & 0.428 \\
	\hline
	$\hat y^{(j)}$ & (0.664, 0.537, 0.501, 0.740) & (0.650, 0.521, 0.485, 0.727) & (0.637, 0.507, 0.470, 0.715) & (0.624, 0.493, 0.455, 0.704) & (0.612, 0.479, 0.442, 0.693) \\
	\hline
	$\mathsf{loss}^{(j)}$ & 0.026 & 0.022 & 0.018 & 0.015 & 0.012 \\
	\hline
\end{tabular}
\end{table}

\subsection{Finding input/output instances based on parameter evolution}
Suppose we have a trace of parameters over several epochs ($w^{(j)}$ and $b^{(j)}$ for $j=0,1,\ldots$), and want to find out the vectors $\mathbf{x}$ and $\mathbf{y}$. The size of vectors $n$ and the learning rate $\eta$ are publicly known. Therefore, we have $2n$ unknowns.

Each new epoch gives us 2 new equations:
\begin{align}
	\pdv{\mse}{w^{(j)}} &= (w^{(j)} - w^{(j+1)}) / \eta  \\
	\pdv{\mse}{b^{(j)}} &= (b^{(j)} - b^{(j+1)}) / \eta
\end{align}
Therefore, we need at least $n$ epochs (besides the initial epoch 0) in order for the system to be solvable.%
\footnote{The term \emph{solvable} is not precise. If the system were linear, we would require $m$ \emph{independent} equations to solve uniquely for $m$ variables. This does not (readily) generalize to non-linear systems.}

Let us rewrite the above equations by expanding the partial derivatives.
\begin{multline}
\begin{cases}
	\sum_{i=0}^{n-1} x_i\Big(\tanh\left(w^{(j)}x_i + b^{(j)}\right) - y_i\Big)\Big(1 - \tanh\left(w^{(j)}x_i + b^{(j)}\right)^2\Big) &= \frac{n}{2\eta}\left(w^{(j)} - w^{(j+1)}\right) \\
	\sum_{i=0}^{n-1} \Big(\tanh\left(w^{(j)}x_i + b^{(j)}\right) - y_i\Big)\Big(1 - \tanh\left(w^{(j)}x_i + b^{(j)}\right)^2\Big) &= \frac{n}{2\eta}\left(b^{(j)} - b^{(j+1)}\right)
\end{cases}\cr
\text{for $j=0,1,\ldots,n$}
\end{multline}
It can be seen that the system is highly non-linear. In general, solving a non-linear system (even if it is only quadratic) is NP-complete in the worst case \cite{blum2000theory}.

\paragraph{Example 1.}
For \Cref{tbl:1}, define $T(x) = \tanh(0.5x + 0.5)$ and $Z(x,y) = (T - y)(1 - T^2)$. Then:
\begin{equation}
\begin{cases}
x\cdot Z(x,y) &= \frac{1}{2\times0.1}(0.500 - 0.489) = 0.055 \\
Z(x,y) &= \frac{1}{2\times0.1}(0.500 - 0.482) = 0.090
\end{cases}
\end{equation}
In this case, the system is actually easy to solve: Dividing the equations, we get $x = \frac{0.055}{0.090} \approx 0.6$, and therefore:
\begin{equation}
	Z(0.6, y) = (\tanh(0.8)-y)(1-\tanh^2(0.8)) = 0.090 \quad\Rightarrow\quad y \approx 0.5
\end{equation}
Therefore, we obtained $\mathbf{x}$ and $\mathbf{y}$ to a very good approximation of \Cref{tbl:1}.

\paragraph{Example 2.}
For \Cref{tbl:2}, define $T_0(x_i) = \tanh(0.5x_i + 0.5)$ and $T_1(x_i) = \tanh(0.493x_i + 0.481)$, and $Z_j(x_i,y_i) = (T_j(x_i) - y_i)(1 - T_j(x_i)^2)$ for $i,j \in \{0,1\}$. Then:
\begin{equation}
\begin{cases}
x_0 \cdot Z_0(x_0, y_0) + x_1 \cdot Z_0(x_1, y_1) &= \frac{2}{2\times0.1}(0.500 - 0.493) = 0.070 \\
Z_0(x_0, y_0) + Z_0(x_1, y_1) &= \frac{2}{2\times0.1}(0.500 - 0.481) = 0.190 \\
x_0 \cdot Z_1(x_0, y_0) + x_1 \cdot Z_1(x_1, y_1) &= \frac{2}{2\times0.1}(0.493 - 0.486) = 0.070 \\
Z_1(x_0, y_0) + Z_1(x_1, y_1) &= \frac{2}{2\times0.1}(0.481 - 0.463) = 0.180 \\
\end{cases}
\end{equation}

Contrary to the previous example, this system is highly nonlinear and cannot be solved analytically. However, using numerical methods can be helpful. 
We tried Python's library SciPy (specifically, the methods \texttt{scipy.optimize.fsolve} and \texttt{scipy.optimize.root}). However, they either didn't converge or came up with unsatisfactory solutions. We also tested SageMath \verb|solve|, which didn't find a solution within a reasonable time.
On the other hand, the following Mathematica\textsuperscript{\tiny\textregistered} code:%
\footnote{the coefficients in \Cref{tbl:2} are not precise enough; here, a 7-digit precision is used.}
\begin{tcolorbox}[breakable, enhanced]
\begin{verbatim}
T0[x_] := Tanh[0.5000000*x + 0.5000000]
T1[x_] := Tanh[0.4925472*x + 0.4810773]
Z0[x_, y_] := (T0[x] - y)*(1 - T0[x]^2)
Z1[x_, y_] := (T1[x] - y)*(1 - T1[x]^2)

FindRoot[{
    x0*Z0[x0, y0] + x1*Z0[x1, y1] == 10*(0.5000000 - 0.4925472),
       Z0[x0, y0] +    Z0[x1, y1] == 10*(0.5000000 - 0.4810773), 
    x0*Z1[x0, y0] + x1*Z1[x1, y1] == 10*(0.4925472 - 0.4855530), 
       Z1[x0, y0] +    Z1[x1, y1] == 10*(0.4810773 - 0.4634884)
    },
    {{x0, 0.5}, {x1, 0}, {y0, 0}, {y1, 0}}
]
\end{verbatim}
\end{tcolorbox}
\noindent returns the following results:
\begin{tcolorbox}[breakable, enhanced]
\begin{verbatim}
{x0 -> 0.599961, x1 -> 0.200118, y0 -> 0.500027, y1 -> 0.400007}
\end{verbatim}
\end{tcolorbox}
\noindent Therefore, we obtained $\mathbf{x}$ and $\mathbf{y}$ to a very good approximation of \Cref{tbl:2}.

\paragraph{Open Question 1.} Are there any result on how to numerically solve this type of system (for large $n$)? Is there any known result stating that solving them are hard \emph{on the average}?

\section{Generalizations}		\label{sec:general}
\subsection{Neural networks with arbitrary layers and nodes}

\begin{figure}[H]
\centering
\includegraphics[width=10cm]{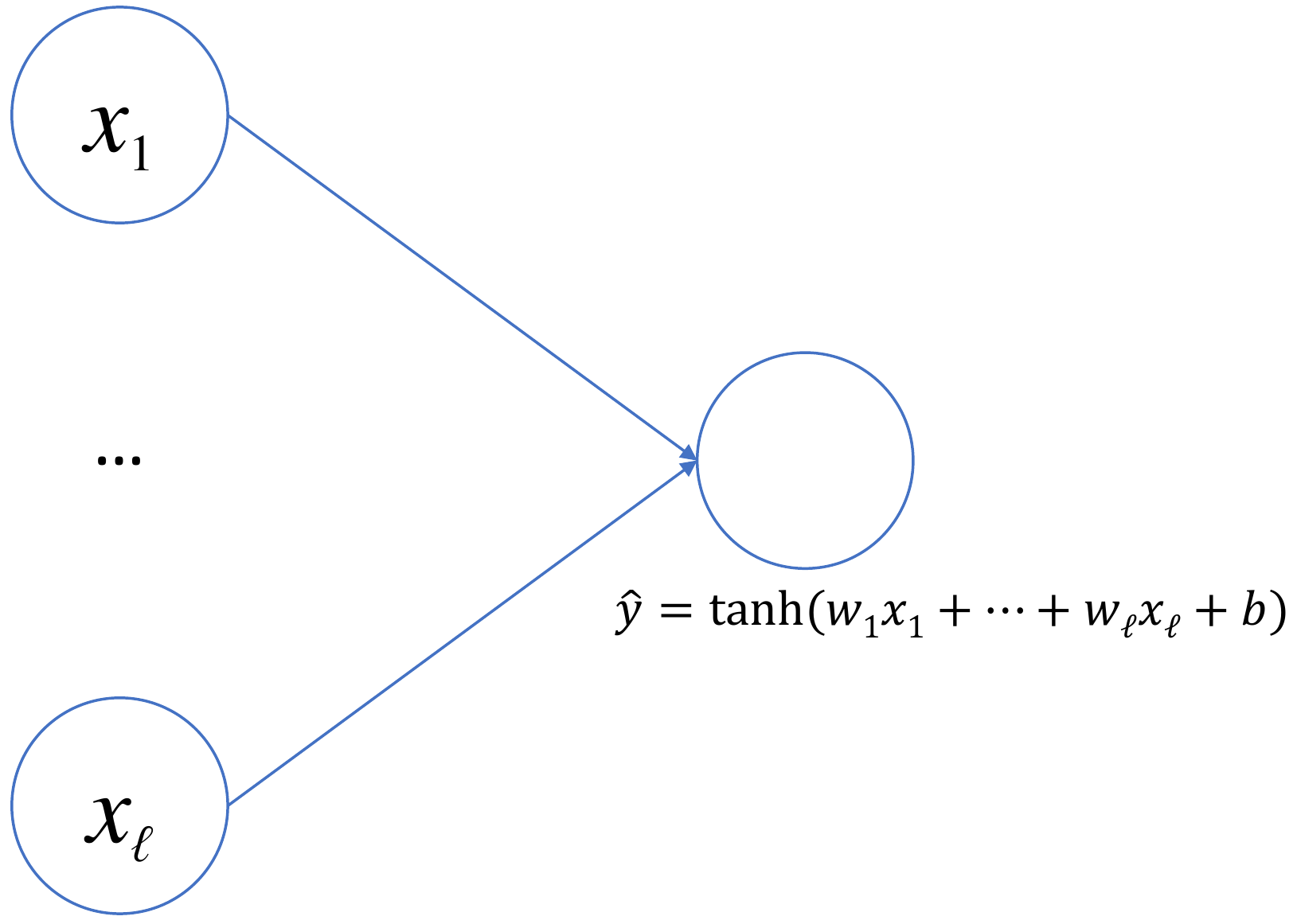}
\caption{A node with fan-in $\ell$.}
\label{fig:nn2}
\end{figure}

In general, a (feed-forward) neural network has many nodes and layers. \Cref{fig:nn2} shows how $\ell$ nodes can contribute to the value of a node in the next layer. Suppose a neural network has $L$ layers (including input and output), and each layer has $\ell$ nodes. Furthermore, assume that the layers are \emph{fully connected}.
Then we have $N = \ell \times L$ nodes and $C = \ell^2 (L-1)$ connections. For $I$ instances and $E$ epochs, we get $N\times I = \ell \times L \times I$ unknowns and $\ell(\ell + 1)(L-1) \times E$ equations.%
\footnote{Each node gives rise to $\ell$ equations for each weight, plus one equation for the bias.}
Therefore, we need $\ell(\ell + 1)(L-1) \times E \ge \ell \times L \times I$ in order for the system to be solvable. Roughly speaking, this means that the number of epochs $E$ must be at least $\frac{I}{\ell}$.

\paragraph{Open Question 2.} What are some real-world examples for $E$, $I$ and $\ell$? Is the above bound satisfiable in practice?

\subsection{Neural networks with randomized learning}
Instead of Gradient Descent (GD), one may opt to use the Stochastic Gradient Descent (SGD). In the latter, instead of using all instances in each epoch, a random subset of instances is used. Therefore, it seems infeasible to fix a system of equations to solve for the inputs.

\paragraph{Open Question 3.} Is it possible to generalize our method to solve the inputs for ANNs with randomized learning?

\section{Conclusion}	\label{sec:conclusion}
In this paper, we discussed a new problem: Learning the input dataset to an artificial neural network, given the network parameters in each epoch. Solving this problem allows an adversary to ruin the privacy of participating parties in a multiparty ANN learning protocol, which leaks the intermediate parameters. We investigated the exact version of the problem, where the adversary is interested in finding the exact dataset, rather than gaining partial knowledge about it.

We showed how the adversary can form a system of equations based on the intermediate values of parameters in each epoch. The system, however, is highly non-linear, and is not easy to solve; especially when the number of equations and unknowns increases. It seems also infeasible to even form the system if the underlying ANN uses a randomized learning approach.

We posed three open questions, and hope that this preliminary investigation brings this type of problem to the attention of the research community. While we considered the problems in the context of a real-world protocol, solving non-linear systems of this type might be of independent theoretical interest.

\bibliographystyle{alpha}
\bibliography{refs}

\appendix
\section{Appendix: Codes}	\label{sec:apx}
This appendix gives the Python code used to generate \Cref{tbl:1,tbl:2,tbl:3,tbl:4}. Variable \texttt{n} determines which table to print. The output precision is set by \verb|np.set_printoptions()|. For instance, you can change \verb|"%.2f"| to \verb|"%.7f"| to increase precision from 2 to 7.

\begin{tcolorbox}[breakable, enhanced]
\begin{verbatim}
import numpy as np

np.set_printoptions(formatter={"float_kind": lambda t: "%.2f" % t})

# how many instances do we need?
n = 4

# x is the input, y is the associated class
x = np.array([0.6, 0.2, 0.1, 0.9][:n])
y = np.array([0.5, 0.4, 0.3, 0.6][:n])

# initial weight and bias
w = 0.5
b = 0.5


# computes a simple neural network:
# 1. a linear transformation
# 2. a Tanh activation function
def nn(): return np.tanh(w * x + b)


# computes the loss function between (MSE)
# real (y) and computed (y_hat) classes
def loss(y_hat):
    z = (np.square(y_hat - y)).mean()
    return z


# computes the partial derivatives of loss function
# with respect to variables w and b
def diffs(y_hat):
    base = 2 * (y_hat - y) * (1 - y_hat ** 2)
    db = np.mean(base)
    dw = np.mean(x * base)
    return dw, db


def main():
    global w, b

    # learning rate
    lr = 0.1

    # number of epochs
    epoch = 5

    W = []
    B = []
    YHAT = []
    LOSS = []

    for _ in range(epoch):
        y_hat = nn()

        W.append(w)
        B.append(b)
        YHAT.append(y_hat)
        LOSS.append(loss(y_hat))

        dw, db = diffs(y_hat)
        w -= lr * dw
        b -= lr * db

    print('w =', np.array(W))
    print('b =', np.array(B))
    print('y_hat =', np.array(YHAT))
    print('loss =', np.array(LOSS))


if __name__ == '__main__':
    main()
\end{verbatim}
\end{tcolorbox}

\end{document}